\documentclass[12pt]{article}

\begin{document}

\title{3-enumerated alternating sign matrices}
\author{Yu.~G.~Stroganov\\
\small \it Institute for High Energy Physics\\[-.5em]
\small \it 142284 Protvino, Moscow region, Russia}
\date{}

\maketitle

\begin{abstract}
Let $A(n,r;3)$ be the total weight of the alternating sign matrices 
of order $n$ whose sole `1' of the first row is at the $r^{th}$ column and  the 
weight of an individual matrix is $3^k$ if it has $k$ entries equal to -1.
Define the sequence of the generating functions $G_n(t)=\sum_{r=1}^n A(n,r;3)\>t^{r-1}$.

Results of two different kind are obtained.
On the one hand I made the explicit expression for the even subsequence  $G_{2\nu}(t)$
in terms of two linear homogeneous second order recurrence in $\nu$ (Theorem 1).
On the other hand I brought to light  the nice connection between the neighbouring functions
$G_{2\nu+1}(t)$ and $G_{2\nu}(t)$ (Theorem 2). 

The  3-enumeration $A(n;3) \equiv G_n(1)$ which was found by Kuperberg is
reproduced as well.
\end{abstract} 

\begin{section}{Introduction and results}

An alternating sign matrix, or ASM, is a matrix of 0's,1's, and -1's such that the non-zero
elements in each row and column alternate between 1 and -1 and 
begin and end with 1.
Recall the conventional notations:
\begin{itemize}
\item
$A(n)$ $\quad$ the number of $n \times n$  ASMs;
\item
$A(n;3)$ $\quad$ the total weight of all $n \times n$  ASMs, where the 
weight of an individual matrix is $3^k$ if it has $k$ entries equal to -1;
\item
$A(n,r)$ $\quad$ the number $n \times n$  ASMs for which the (unique) `1' of the first row is
at
the
$r^{th}$
column;
\item
$A(n,r;3)$ $\quad$  the total weight of $n \times n$  ASMs for which the 
(unique) `1' of the first row is at the $r^{th}$ column and  the 
weight of an individual matrix is $3^k$ if it has $k$ entries equal to -1;
\end{itemize}
An ultra-short background can be described by the scheme:

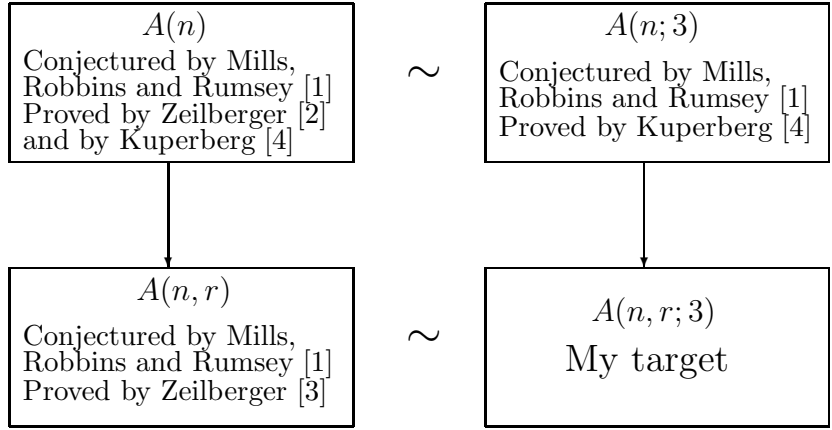
\begin{picture}(0,190)

\put(20,20){\line(1,0){130}}
\put(20,80){\line(1,0){130}}
\put(20,120){\line(1,0){130}}
\put(20,180){\line(1,0){130}}

\put(200,20){\line(1,0){130}}
\put(200,80){\line(1,0){130}}
\put(200,120){\line(1,0){130}}
\put(200,180){\line(1,0){130}}

\put(20,20){\line(0,1){60}}
\put(20,120){\line(0,1){60}}

\put(150,20){\line(0,1){60}}
\put(150,120){\line(0,1){60}}

\put(200,20){\line(0,1){60}}
\put(200,120){\line(0,1){60}}

\put(330,20){\line(0,1){60}}
\put(330,120){\line(0,1){60}}

\put(70,168){$A(n)$}
\put(25,155){\small Conjectured by Mills,}
\put(25,145){\small Robbins and Rumsey \cite{MRR}}
\put(25,135){\small Proved by Zeilberger \cite{Z0}}
\put(25,125){\small and by Kuperberg \cite{Ku1}}

\put(245,168){$A(n;3)$}
\put(205,151){\small Conjectured by Mills,}
\put(205,141){\small Robbins and Rumsey \cite{MRR}}
\put(205,130){\small Proved by Kuperberg \cite{Ku1}}

\put(68,68){$A(n,r)$}
\put(25,51){\small Conjectured by Mills,}
\put(25,41){\small Robbins and Rumsey \cite{MRR}}
\put(25,30){\small Proved by Zeilberger \cite{Z}}

\put(240,60){$A(n,r;3)$}
\put(230,41){\large My target}

\put(80,120){\vector(0,-1){40}}
\put(260,120){\vector(0,-1){40}}
\put(170,150){\Large $\sim$}
\put(170,50){\Large $\sim$}
\end{picture}

We investigate the generating function 
$G_{n}(t)=\sum_{r=1}^{n} A(n,r;3)\>t^{r-1}$.
The next results follow directly from the general method elaborated in paper \cite{my1}.

\bf{Theorem 1} \rm (ASMs of the even order)
The normalized generating function 
$\tilde{G}_{2\nu+2}(t)\equiv G_{2\nu+2}(t)/A(2\nu+2;3)$ is given
by
a formula
\begin{eqnarray}
&&\tilde{G}_{2\nu+2}(t)=\frac{(2\nu+1)!}{(3\nu+2)!}\>
 \frac{\{(3\nu+2)g^{(1)}_{\nu}(t)-(3\nu+1)g^{(2)}_{\nu}(t)\}}{t+1},\quad \nu \ge 0 \nonumber
\end{eqnarray}
where $g^{(j)}_{\nu}(t), \quad (j=1,2)$ are two sequences of the polynomials which are fixed
by the second order recurrence
\begin{eqnarray}
&&3\>(1-t^2)^2\>g^{(j)}_{\nu+1}(t)-\nonumber \\
&&-(2\nu+1)\>(1+4t+t^2)\>[3(1+t+t^2)^2-(1+4t+t^2)^2]\>g^{(j)}_{\nu}(t)-\nonumber \\
&&-(9\>\nu^2-j^2)\>t^2\>(1+2t)^2\>(2+t)^2\>g^{(j)}_{\nu-1}(t)=0, \quad
(j=1,2)\nonumber
\end{eqnarray} 
with the initial data
\begin{eqnarray}
&&g^{(1)}_0(t)=(t^2+t+1)/3, \quad g^{(1)}_1(t)=[(1+2t)^2(2+t)^2-9t^2]/18 \nonumber\\
&&g^{(2)}_0(t)=-(t^2+4t+1)/3,\>\> g^{(2)}_1(t)=-[(1+2t)^2(2+t)^2+9t^2]/18.\nonumber
\end{eqnarray} 

\bf{Theorem 2} \rm 
The normalized generating functions $\tilde{G}_{n}(t)\equiv G_{n}(t)/A(n;3)$ satisfy 
\begin{eqnarray}
&&\tilde{G}_{2\nu+1}(t)=\tilde{G}_{2\nu}(t)\frac{2\>(1+2t)(2+t)}{9\>(t+1
)},
\quad \nu > 0 \nonumber
\end{eqnarray}

As far as the normalizing factor $A(n;3)$ is  concerned one can easily reproduce
Theorem 2 of Kuperberg paper~\cite{Ku1} using equations of Section 4. 
\end{section}

\begin{section}{Outline of method}

This small note can be considered as the supplement to my recent paper \cite{my1}.  I use the
notation and results of the cited paper without reviewing them.

Consider the inhomogeneous six-vertex model with the domain wall boundary conditions
\cite{kor1}.
 Spectral parameters $\{x_1,x_2,...,x_{n}\}$ and $\{y_1,y_2,...,y_{n}\}$ are attached to the
 horizontal and vertical lines respectively. We  fix them as follows:
\begin{eqnarray}
\label{case}
&& x_1=u ,\nonumber \\
&& x_i=\pi/2,\quad i=2,3,...,n, \nonumber \\
&& y_i=0,\quad i=1,2,...,n. \nonumber
\end{eqnarray}
The 3-enumeration of ASMs corresponds to the 6-vertex model with the special crossing
parameter
$\eta=2\pi/3$, so that  the Boltzmann weights in the  upper row are given by:
 \begin{eqnarray}
\label{upperBW}
&&a(u)=2 \sin (u+\pi/3), \quad b(u)=2 \sin (u-\pi/3),\\
 &&c(u)=\sqrt{3} \nonumber.
\end{eqnarray}
All weights in the remaining rows\footnote{We use formulae (1) of paper~\cite{my1} changing
the
normalization by a factor $\sqrt{3}$.} correspond to the 3-enumeration:
\begin{eqnarray}
\label{otherBW}
&&a=b=1 \quad c=\sqrt{3}.\nonumber
\end{eqnarray}
The partition function of the model is given by the trigonometric polynomial of degree $n-1$:
\begin{eqnarray}
\label{1stZ}
&&Z(u)=\sqrt {3} \sum_{r=1}^n A(n,r;3)\> a^{r-1}(u)\> b^{n-r}(u).
\end{eqnarray}
Using equations (\ref{upperBW}), (\ref{1stZ}) and right-left symmetry of ASMs  we obtain 
that $Z(u)$  satisfy
\begin{eqnarray}
\label{pZ}
&&Z(-u)=(-1)^{n-1} Z(u).
\end{eqnarray}
It was shown in paper~\cite{my1} (see equations (5-7) of the cited paper)
that one has the equation 
\begin{eqnarray}
 \label{eq15}
&&f(u)+f(u+\frac{2\pi}{3})+f(u+\frac{4\pi}{3})=0 
\end{eqnarray}
with
\begin{eqnarray}
 \label{eq14}
&&f(u) = Z(u)\>\sin^n(u)\>\cos^{n-1}(u).
\end{eqnarray}
We see that the function $f(u)$ is an odd trigonometric polynomial of degree
$3n-2$.
It turns out that under the circumstances the function $f(u)$ and consequently the partition
function $Z(u)$ can be found up to an arbitrary constant.
Solving these equations (the solution is given in Appendix A) we obtain
\begin{eqnarray}
\label{evenf}
&&f_{2\nu+2}(u)=c_{\nu}\nonumber \\
&&\biggl\{ (3\nu+2) \sum_{\alpha =0}^{\nu}
\biggl(\begin{array}{c}
\nu - \frac{1}{3} \\
\alpha
\end{array}\biggr)
\biggl(\begin{array}{c}
\nu + \frac{1}{3} \\
\nu - \alpha
\end{array}\biggr)
\sin(2-6\nu+12\alpha) u - \\
&&-(3 \nu +1) \sum_{\alpha =0}^{\nu}
\biggl(\begin{array}{c}
\nu - \frac{2}{3} \\
\alpha
\end{array}\biggr)
\biggl(\begin{array}{c}
\nu + \frac{2}{3} \\
\nu - \alpha
\end{array}\biggr)
\sin(4-6\nu+12\alpha) u \biggr\}\nonumber
\end{eqnarray} 
for $n$ even and
\begin{eqnarray}
\label{oddf}
&&f_{2\nu+1}(u)=r_{\nu}\cos 3u\>f_{2\nu}(u)
\end{eqnarray} 
for $n$ odd,
where multipliers $c_{\nu}$ and $r_{\nu}$ do not depend on $u$. 
$\nu$ equals to the integer part of $(n-1)/2$.

Formulae (\ref{upperBW}), (\ref{1stZ}), (\ref{eq14}), (\ref{evenf}) and (\ref{oddf}) solve in
a
sense 
the problem. We are left with the task of elaborating an effective method for 
explicit calculation of $A(n,r;3)$. Besides we have to find the multipliers $c_{\nu}$
and $r_{\nu}$. 
\end{section}

\begin{section}{Recursion relations}
Let us begin with $n$ even.
It is convenient to present equation (\ref{evenf}) as follows
\begin{eqnarray}
\label{evenfs}
&&f_{2\nu+2}(u)=c_{\nu}
[(3\nu+2)\>F^{(1)}_{\nu}(u)-(3\nu+1)\>F^{(2)}_{\nu}(u)]
\end{eqnarray} 
where $F^{(j)}_{\nu}(u)$, $(j=1,2)$ are trigonometric polynomials
\begin{eqnarray}
\label{Fc}
&&F^{(j)}_{\nu}(u)=\sum_{\alpha =0}^{\nu}
\biggl(\begin{array}{c}
\nu - \frac{j}{3} \\
\alpha
\end{array}\biggr)
\biggl(\begin{array}{c}
\nu + \frac{j}{3} \\
\nu - \alpha
\end{array}\biggr)
\sin 2\>(j-3\nu+6\alpha) u \nonumber
\end{eqnarray}

One can verify directly that these polynomials satisfy a recurrence
\begin{eqnarray}
\label{recF}
&&9\nu \> (\nu+1)\>F^{(j)}_{\nu+1}(u)-18\>\nu\>(2\nu+1)\>\cos 6u\>
F^{(j)}_{\nu}(u)-\nonumber \\
&&-4\>(9\>\nu^2-j^2)\>\sin^2 6u\> F^{(j)}_{\nu-1}(u)=0, \quad (j=1,2)
\end{eqnarray} 
with the initial data
\begin{eqnarray}
\label{ini}
&&F^{(1)}_0(u)=\sin 2u, \quad F^{(1)}_1(u)=\frac{2}{3}(\sin 8u-2\sin 4u) \nonumber \\ 
&&F^{(2)}_0(u)=\sin 2u, \quad F^{(2)}_1(u)=\frac{1}{3}(\sin 10u-5\sin 2u) \nonumber
\end{eqnarray} 

It was shown in paper~\cite{odd} that quotients
\begin{eqnarray}
\label{quo}
&&\Phi^{(j)}_{\nu}(w) = F^{(j)}_{\nu}(u)/\sin^{2\nu+1} 2u,\quad
(j=1,2) 
\end{eqnarray} 
are polynomials in the variable $w = \cos 2u$.
Recurrence (\ref{recF}) can be rewritten as a recurrence for
these polynomials:
\begin{eqnarray}
\label{recPhi}
&&9\nu \> (\nu+1)\>(1-w^2)\>\Phi^{(j)}_{\nu+1}(w)+18\>\nu\>(2\nu+1)\>w\>(3-4w^2)\>
\Phi^{(j)}_{\nu}(w)-\nonumber
\\
&&-4\>(9\>\nu^2-j^2)\>(1-4w^2)^2\>\Phi^{(j)}_{\nu-1}(w)=0, \quad (j=1,2)
\end{eqnarray} 
with the initial data
\begin{eqnarray}
\label{iniPhi}
&&\Phi^{(j)}_0(w;1)=1, \quad \Phi^{(j)}_1(w;1)=-16\>w/3 \nonumber \\
&&\Phi^{(j)}_0(w;2)=2\>w, \quad \Phi^{(j)}_1(w;2)=-4\>(1+4w^2)/3
\end{eqnarray} 

Combining equations (\ref{1stZ}), (\ref{eq14}), (\ref{evenfs}) and (\ref{quo})
we obtain firstly
\begin{eqnarray}
\label{2ndZp}
&&Z(u)\sin u\equiv \sqrt {3}\sin u\> a^{2\nu+1}(u) \sum_{r=1}^{2\nu+2} A(2\nu+2,r;3)\> 
\biggl(\frac {b(u)}
{a(u)}\biggr)^{2\nu+2-r}=\nonumber \\
&&=2^{2\nu}\>c_{\nu}\>\{(3\nu+2)\>\Phi^{(1)}(w)-(3\nu+1)\>\Phi^{(2)}(w)\},\nonumber
\end{eqnarray}
and using a new variable $t$
\begin{eqnarray}
\label{tvar}
&&t=\frac {b(u)}{a(u)}=\frac{\sin(u-\pi/3)}{\sin(u+\pi/3)},
\end{eqnarray}
we can present the last equation as follows
\begin{eqnarray}
\label{Geven}
&&G_{2\nu+2}(t)\equiv  \sum_{r=1}^{2\nu+2} A(2\nu+2,r;3)\>t^{2\nu+2-r}= \nonumber \\
&&=\biggl(\frac{4}{3}\biggr)^{\nu+1}\frac{c_{\nu}}{\sqrt{3}\>(t+1)}
(t^2+t+1)^{\nu+1}\\
&&\biggl\{(3\nu+2)\> \Phi^{(1)}_{\nu}(-\frac{t^2+4t+1}{2(t^2+t+1)})-
(3\nu+1)\> \Phi^{(2)}_{\nu}(-\frac{t^2+4t+1}{2(t^2+t+1)})
\biggr\} \nonumber
\end{eqnarray}
where  $c_{\nu}$ is a multiplier  which does not depend
on $t$ (see equation (\ref{evenf})).

It is clear that $A(2\nu+2,r;3)=A(2\nu+2,2\nu+3-r;3)$ therefore we can write
$$G_{2\nu+2}(t)\equiv  \sum_{r=1}^{2\nu+2} A(2\nu+2,r;3)\>t^{r-1}$$
in the first part of equation
(\ref{Geven}).

On the other hand combining equations (\ref{1stZ}), (\ref{eq14}) and (\ref{oddf})
we obtain firstly the relation between the partition functions:
\begin{eqnarray}
\label{ZZ}
&&Z_{2\nu+1}(u)\sin u \cos u= r_{\nu} \cos 3u\>Z_{2\nu}(u),\nonumber
\end{eqnarray}
and then transform it
into the equation connecting the generating functions
\begin{eqnarray}
\label{GG}
&&G_{2\nu+1}\biggl(\frac{b(u)}{a(u)}\biggr) a(u)\sin u = r_{\nu} \frac{\cos 3u}{\cos u}
G_{2\nu}\biggl(\frac{b(u)}{a(u)}\biggr).\nonumber
\end{eqnarray}
Using the variable $t$ (see equation (\ref{tvar}) we rewrite the last equation as 
\begin{eqnarray}
\label{GGt}
&&G_{2\nu+1}(t) = -r_{\nu} \frac{2(1+2t)(2+t)}{3(1+t)}
G_{2\nu}(t)
\end{eqnarray}
where  $r_{\nu}$ is a multiplier  which does not depend
on $t$.

\end{section}

\begin{section}{Multipliers $c_{\nu}$ and $r_{\nu}$} 

We are prepared now to calculate the multipliers $c_{\nu}$ and $r_{\nu}$.
The point that for some special values of the spectral 
parameter $u$ (see equation (\ref{case})) the recurrences we consider are simplified
and can be easily solved. Details can be found in Appendix B.
Let us compare equations (\ref{Aeven}), (\ref{Aodd}), (\ref{Aeven2}) and (\ref{Aodd2}).
Excluding the multipliers from these equations  we obtain

\begin{eqnarray}
\label{Aratio1}
&&\frac{A(2\nu+2;3)}{A(2\nu+1;3)}=\frac{3^{\nu} \nu ! (3\nu+2)!}{[(2\nu+1)!]^2} \nonumber
\end{eqnarray} 
and
\begin{eqnarray}
\label{Aratio2}
&&\frac{A(2\nu+1;3)}{A(2\nu;3)}=\frac{9A(2\nu;3)}{4A(2\nu-1;3)}=\frac{3^{\nu} \nu !
(3\nu)!}{[(2\nu)!]^2}. \nonumber
\end{eqnarray} 
Using these fractions and the trivial equation $A(1;3)=1$ one can reproduce Theorem 2 of 
Kuperberg paper~\cite{Ku1}. Fixing the multipliers $c_{\nu}$ from equation (\ref{Aeven2}),
inserting them into (\ref{Geven}) and defining new polynomials:
\begin{eqnarray}
\label{Lastg}
&&g^{(j)}_{\nu}(t)=\frac{\nu ! (t^2+t+1)^{\nu+1}}{3\>4^\nu}
\Phi^{(j)}_{\nu}\biggl(-\frac{t^2+4t+1}{2(t^2+t+1)}\biggr) \nonumber
\end{eqnarray} 
we obtain our Theorem 1.
On the other hand we easily obtain the Theorem 2 normalizing the generation function
entering equation (\ref{GGt}).

\vspace{0.5cm}

\bf {\it Acknowledgments} \rm The work was supported in part by the
Russian Foundation for Basic Research under grant \# 01--01--00201
and by the INTAS under grant \# 00--00561.

\end{section}

\vspace{0.5cm}

\bf{Appendix A. Calculation of $f(u)$ up to arbitrary constant} \rm

Recall that $f(u)$ is an odd trigonometric polynomial of degree $3n-2$.   
We can therefore write
\begin{eqnarray}
\label{Fur}
&&f(u)=b_1 \sin (3n-2) u +b_2 \sin (3n-4) u +b_3 \sin (3n-6) u +...
\end{eqnarray}
Equation (\ref{eq15}) is satisfied if and only if $b_{3 \kappa}=0$.
This condition implies
\begin{eqnarray}
\label{export}
&&f(u)=\sum_{k=0}^{n-1} \beta _{k} \sin (4-3n+6k)u. 
\end{eqnarray}
According to equation (\ref{eq14}) the monomial $\sin^n u\>\cos^{n-1} u$ divides this
function and
we obtain a set of equations 
\begin{eqnarray}
\label{set}
&&f^{(m)}(u)|_{u=0}=0 \quad m=0,1,...,n-1,\nonumber \\
&&f^{(m)}(u)|_{u=\pi /2}=0 \quad m=0,1,...,n-2,
\end{eqnarray}
 where $f^{(m)}$ the derivative of $f$ of the order $m$.
Let us insert equation (\ref{export}) into (\ref{set}).

If $n$ is even then we obtain the system:
\begin{eqnarray}
\label{even1}
&&\sum_{k=0}^{2\nu+1} \beta _{k} (3k-1-3\nu)^{2\mu +1}=0,\quad \mu=0,1,...,\nu \nonumber \\
&& \sum_{k=0}^{2\nu+1} (-1)^k\> \beta _{k} (3k-1-3\nu)^{2\mu +1}=0,\quad \mu=0,1,...,\nu-1.
\nonumber
\end{eqnarray}
Considering separately odd and even $k$ and combining obtained relations
we get two similar systems 
\begin{eqnarray}
\label{even2}
&&\sum_{\kappa=0}^{\nu} \beta _{2 \kappa} (6\kappa -1-3\nu)^{2\mu +1} =0, \quad
\mu=0,1,...,\nu
-1,
\end{eqnarray}
\begin{eqnarray}
\label{even3}
&&\sum_{\kappa=0}^{\nu} \beta _{2 \kappa +1} (6\kappa +2-3\nu)^{2\mu +1}=0, \quad
\mu=0,1,...,\nu
-1,
\end{eqnarray}
and one additional relation:
\begin{eqnarray}
\label{even4}
&&\sum_{\kappa=0}^{\nu} \beta _{2 \kappa} (6\kappa -1-3\nu)^{2\nu +1} +\sum_{\kappa=0}^{\nu}
\beta
_{2 \kappa +1} (6\kappa +2-3\nu)^{2\mu +1}=0.
\end{eqnarray}
Systems (\ref{even2}) and (\ref{even3}) are
equivalent to the condition that the relations 
\begin{eqnarray}
\label{even5}
&&\sum_{\kappa=0}^{\nu} \beta _{2\kappa} (6\kappa -1-3\nu)\>p((6\kappa -1-3\nu)^2)=0,
\end{eqnarray}
and
\begin{eqnarray}
\label{even6}
&&\sum_{\kappa=0}^{\nu} \beta _{2\kappa+1} (6\kappa +2-3\nu)\>p((6\kappa +2-3\nu)^2)=0,
\end{eqnarray}
are valid for all polynomials $p(x)$ of degree $\nu-1$.
Let us begin with system (\ref{even5}).
Consider $\nu$ special polynomials of degree $\nu-1$:
 \begin{eqnarray}
&&p_{\alpha}(x)=\prod_{\kappa=1,\kappa \ne \alpha}^{\nu} (x-(6\kappa -1-3\nu)^2) \quad
\alpha=1,2,...,\nu. \nonumber
\end{eqnarray}
Inserting these polynomials into (\ref{even5}) we find simple relations connecting $\beta
_{2\alpha}$ with 
$\beta _0$ which can be written as 
\begin{eqnarray}
\label{b2a}
&&\beta_{2\alpha}=c_1\>
\biggl(\begin{array}{c}
\nu + \frac{1}{3} \\
\alpha
\end{array}\biggr)
\biggl(\begin{array}{c}
\nu - \frac{1}{3} \\
\nu - \alpha
\end{array}\biggr)
\quad \alpha =0,...,\nu.
\end{eqnarray} 
In the same way we convert equation (\ref{even6}) into
\begin{eqnarray}
\label{b2a1}
&&\beta_{2\alpha+1}=c_2\>
\biggl(\begin{array}{c}
\nu - \frac{2}{3} \\
\alpha
\end{array}\biggr)
\biggl(\begin{array}{c}
\nu + \frac{2}{3} \\
\nu - \alpha
\end{array}\biggr)
\quad \alpha =0,...,\nu.
\end{eqnarray} 
We have to satisfy the additional relation (\ref{even4}).
Taking into account equations (\ref{b2a}) and (\ref{b2a1}) one can reduce this relation 
to the equation:
\begin{eqnarray}
\label{even4n}
&&c_2/c_1=-\frac{\sum_{\alpha=0}^{\nu}
\biggl(\begin{array}{c}
\nu + \frac{1}{3} \\
\alpha
\end{array}\biggr)
\biggl(\begin{array}{c}
\nu - \frac{1}{3} \\
\nu - \alpha
\end{array}\biggr)
(6\alpha -1-3\nu)^{2\nu +1}}{\sum_{\alpha=0}^{\nu}
\biggl(\begin{array}{c}
\nu - \frac{2}{3} \\
\alpha
\end{array}\biggr)
\biggl(\begin{array}{c}
\nu + \frac{2}{3} \\
\nu - \alpha
\end{array}\biggr) (6\alpha +2-3\nu)^{2\mu +1}}.\nonumber
\end{eqnarray}
Using Mathematica I found that the right side fraction probably equals to 
$(3\nu+1)/(3\nu+2)$. Then I appealed to the domino forum and a few days a proof of 
more general identity was given by Guoce Xin~\cite{Xin}. A simpler proof of Guoce Xin's 
identity was presented by Ira Gessel~\cite{Ira}.

As a result we obtain (up to an arbitrary constant multiplier):
\begin{eqnarray}
\label{feven}
&&f(u)\equiv f_{2\nu+2}(u) \propto \nonumber \\
&&\sim \biggl\{(3\nu+2)\sum_{\alpha =0}^{\nu}
\biggl(\begin{array}{c}
\nu + \frac{1}{3} \\
\alpha
\end{array}\biggr)
\biggl(\begin{array}{c}
\nu - \frac{1}{3} \\
\nu - \alpha
\end{array}\biggr)
\sin(-2-6\nu+12\alpha) u + \nonumber \\
&&+(3\nu+1)\sum_{\alpha =0}^{\nu}
\biggl(\begin{array}{c}
\nu - \frac{2}{3} \\
\alpha
\end{array}\biggr)
\biggl(\begin{array}{c}
\nu + \frac{2}{3} \\
\nu - \alpha
\end{array}\biggr)
\sin(4-6\nu+12\alpha) u\biggr\}.
\end{eqnarray} 
 
Changing $\alpha$ with $\nu-\alpha$ in the first sum we obtain a variation of this formula
which
is used in the body of the paper. 

If $n$ is odd then inserting (\ref{export}) into (\ref{set}) we obtain
\begin{eqnarray}
\label{odd1}
&&\sum_{k=0}^{2\nu} \beta _{k} (6k+1-6\nu)^{2\mu +1}=0,\quad\sum_{k=0}^{2\nu} (-1)^k\> \beta
_{k}
(6k+1-6\nu)^{2\mu}=0,\nonumber \\
 &&\mu=0,1,...,\nu-1, \nonumber
\end{eqnarray}
 where
$n=2 \nu +1$.
Consider separately  $k$ odd and even:
\begin{eqnarray}
\label{odd2}
&&\sum_{\kappa=0}^{\nu} \beta _{2 \kappa} (12\kappa +1-6\nu)^{2\mu +1} +
\sum_{\kappa=1}^{\nu}
\beta _{2 \kappa +1} (12\kappa -5 -6\nu)^{2\mu +1}=0, \nonumber \\
&&\sum_{\kappa=0}^{\nu} \beta _{2\kappa} (12\kappa +1-6\nu)^{2\mu}- \sum_{\kappa=1}^{\nu}
\beta
_{2\kappa +1} (12\kappa -5 -6\nu)^{2\mu}=0,\nonumber  \\
&& \mu=0,1,...,\nu -1. \nonumber
\end{eqnarray}
This system is equivalent to the condition that the relation 
\begin{eqnarray}
\label{odd3}
&&\sum_{\kappa=0}^{\nu} \beta _{2\kappa} p(12\kappa +1 -6\nu)- 
\sum_{\kappa=1}^{\nu} \beta _{2\kappa +1} p(-12\kappa +5 +6\nu)=0
\end{eqnarray}
is valid for all polynomials $p(x)$ of degree $2\nu-1$.
Consider $2\nu$ special polynomials of degree $2\nu-1$:
 \begin{eqnarray}
&&p_{\alpha}(x)=\prod_{\kappa=1,\kappa \ne \alpha}^{\nu} (x-1+6\nu-12\kappa)
\prod_{\kappa=1}^{\nu} (x-5-6\nu+12\kappa) \nonumber \\
&&\tilde p_{\alpha}(x)=\prod_{\kappa=1}^{\nu} (x-1+6\nu-12\kappa)
\prod_{\kappa=1,\kappa \ne \alpha}^{\nu} (x-5-6\nu+12\kappa) \nonumber \\
&& \alpha=1,2,...,\nu. \nonumber
\end{eqnarray}
Inserting these polynomials into (\ref{odd3}) we find simple relations connecting $\beta
_{\alpha}$ with 
$\beta _0$. It is possible to write the answer in terms of binomial coefficients:
\begin{eqnarray}
\label{fodd}
&&f(u)\equiv f_{2\nu+1}\propto \nonumber \\
&&\sim \biggl\{\sum_{\alpha =0}^{\nu}
\biggl(\begin{array}{c}
\nu - \frac{2}{3} \\
\alpha
\end{array}\biggr)
\biggl(\begin{array}{c}
\nu - \frac{1}{3} \\
\nu - \alpha
\end{array}\biggr)
\sin(1-6\nu+12\alpha) u - \nonumber \\
&&-\sum_{\alpha =0}^{\nu -1}
\biggl(\begin{array}{c}
\nu - \frac{1}{3} \\
\alpha
\end{array}\biggr)
\biggl(\begin{array}{c}
\nu - \frac{2}{3} \\
\nu - \alpha -1
\end{array}\biggr)
\sin(5-6\nu+12\alpha) u\biggr\} .\nonumber
\end{eqnarray}

One can check directly that multiplying $f_{2\nu+2}$ (see equation (\ref{feven})) by $\cos
3u$
one obtain $f_{2\nu+3}$ which is given by the last equation with $\nu \rightarrow \nu+1$. 

\vspace{0.5cm}

\bf{Appendix B. Generating functions in special points} \rm

Let us begin with the point $u=\pi/3$. For this case variable $t$ which is defined by equation
(\ref{tvar})
equals to zero and the generating function $G_n(t)$ comes to the first term $A(n,1;3)$.
According to the trivial bijection between the ASMs of order $n$ whose sole `1' of the first
row
is at the first column and all ASMs of order $n-1$ one can write $A(n-1;3)$ instead of
$A(n,1;3)$. 
Equation (\ref{recPhi}) reduces to ($w\equiv\cos 2u=-1/2$)
\begin{eqnarray}
\label{recPhi1}
&&3(\nu+1)\>\Phi^{(j)}_{\nu+1}(-1/2)-\>8 (2\nu+1)\>
\Phi^{(j)}_{\nu}(-1/2)=0, \quad (j=1,2)
\end{eqnarray} 
According to equation (\ref{iniPhi}) we have the initial data
\begin{eqnarray}
\label{iniPhi1}
&&\Phi^{(j)}_{\nu+1}(-1/2)=(-1)^{j+1}\frac{8}{3}, \quad (j=1,2)
\end{eqnarray} 
Recurrence (\ref{recPhi1}) with  initial data (\ref{iniPhi1}) gives
\begin{eqnarray}
\label{Phi1}
&&\Phi^{(j)}_{\nu+1}(-1/2)=(-1)^{j+1}\biggl(\frac{4}{3}\biggr)^{\nu}\frac{(2\nu)!}{(\nu!)^2}.
\nonumber
\end{eqnarray} 
and equation (\ref{Geven}) comes to
\begin{eqnarray}
\label{Aeven}
&&A(2\nu+1;3)=\biggl(\frac{4}{3}\biggr)^{2\nu+1}\frac{c_{\nu}\>\sqrt{3}\>(2\nu+1)!}{(\nu
!)^2}.
\end{eqnarray} 

It is clear that equation (\ref{GGt}) reduces to
\begin{eqnarray}
\label{Aodd}
&&A(2\nu;3)=-\frac{4}{3} r_{\nu} A(2\nu-1;3).
\end{eqnarray} 

Consider now another convenient value of the spectral parameter
$u=\pi/2$. Unlike the previous case variable $t$ which is defined by equation (\ref{tvar})
equals to unity  and the generating function $G_n(t)$ come to 3-enumeration of all
ASMs of order $n$:
\begin{eqnarray}
\label{case2}
&&G_n(1)=A(n;3). \nonumber
\end{eqnarray} 

Equation (\ref{recPhi}) reduces to ($w\equiv\cos 2u=-1$)
\begin{eqnarray}
\label{recPhi2}
&&\nu\>(2\nu+1)\>\Phi^{(j)}_{\nu}(-1)-2\>(9\>\nu^2-j^2)\>\Phi^{(j)}_{\nu-1}(-1)=0, \quad
(j=1,2)
\end{eqnarray} 
with the initial data
\begin{eqnarray}
\label{iniPhi2}
&&\Phi^{(1)}_0(-1)=1, \quad \Phi^{(1)}_1(-1)=16/3 \nonumber \\
&&\Phi^{(2)}_0(-1)=-2, \quad \Phi^{(2)}_1(-1)=-20/3.
\end{eqnarray} 
Recurrence (\ref{recPhi2}) with  initial data (\ref{iniPhi2}) gives
\begin{eqnarray}
\label{Phi2}
&&\Phi^{(1)}_{\nu}(-1)=\biggl(\frac{4}{3}\biggr)^{\nu}\frac{(3\nu+1)!}{\nu!\>(2\nu+1)!}
\nonumber \\
&&\Phi^{(2)}_{\nu}(-1)=-\biggl(\frac{4}{3}\biggr)^{\nu}\frac{(3\nu+2)\>(3\nu)!}{\nu!\>(2\nu+1)
!}\nonumber.
\end{eqnarray} 
and equation (\ref{Geven}) comes to
\begin{eqnarray}
\label{Aeven2}
&&A(2\nu+2;3)=\biggl(\frac{16}{3}\biggr)^{\nu}\frac{4\>c_{\nu}\>(3\nu+2)!}
{\sqrt{3}\>\nu !\>(2\nu+1)!}.
\end{eqnarray} 

On the other hand equation (\ref{GGt}) reduces to
\begin{eqnarray}
\label{Aodd2}
&&A(2\nu+1;3)=-3 r_{\nu} A(2\nu;3).
\end{eqnarray}

\end{document}